\documentclass[12pt]{article}
\usepackage{subfigure}
\usepackage{amssymb,amsmath}
\usepackage{graphicx}
\usepackage{float}
\usepackage{color}
\usepackage[font=footnotesize]{caption}
\usepackage{array,multirow}
\usepackage[colorlinks=true
,urlcolor=blue
,citecolor=blue
,linkcolor=blue
,pagecolor=blue
,linktocpage=true
,pdfproducer=medialab
]{hyperref}
\usepackage[a4paper]{geometry}
\usepackage{cite}
\usepackage{placeins}
\usepackage[inline]{enumitem}
\usepackage{cancel}
\usepackage{arydshln}
\makeatletter \renewcommand{\@dotsep}{10000} \makeatother
\usepackage{indentfirst}
\def\beq{\begin{equation}}
\def\eeq{\end{equation}}
\usepackage[nodisplayskipstretch]{setspace}

\begin{document}

\begin{titlepage}

\begin{flushright}

\end{flushright}
\pagestyle{empty}

%\vspace*{0.2in}
\begin{center}
	{\Large \bf Investigating non-minimal flavour-violating CMSSM in the light of Higgs-Boson mass using information theory}\\ 
	\vspace{8pt}
	{\bf  Surabhi Gupta\footnote{\tt E-mail:sgupta2@myamu.ac.in} and Sudhir Kumar Gupta\footnote{\tt E-mail:sudhir.ph@amu.ac.in} }
	\vspace{2pt}
	\begin{flushleft}
		{\em Department of Physics, Aligarh Muslim University, Aligarh, UP--202002, India} 
	\end{flushleft}
	
	\vspace{10pt}
	\begin{abstract}
Flavour-violating interactions of the stop-quarks are expected to provide an additional few GeV contributions to the Higgs-Boson mass, particularly when mix with scharm-quarks, thereby allowing reduced supersymmetry (SUSY) breaking scale compared to flavour-conserving constrained minimal supersymmetric Standard Model (CMSSM). Inspired by this, we analyse the interactions mentioned above in the context of CMSSM using the information entropy of the Higgs-Boson for a wider region of flavour-violating CMSSM parameter space $(m_0,m_{1/2},A_0,$ $\tan\beta,sgn(\mu),\delta_{ct}^{ij})$, where $\delta_{ct}^{ij}$ represents the flavour-violating coupling of the top-quark with the charm-quark and $i, j$ defining left and right chiralities of squarks. Our information-theoretic analysis of the model mentioned above reveals the most favourable values of $(m_0, m_{1/2},A_0, \tan\beta,\delta_{ct}^{ij})$ as $(4.30 {\rm~TeV}, 2.32 {\rm~TeV}, -4.96{\rm~ TeV},$ $ 22.8,0.037)$ and $(4.16 {\rm ~TeV},$ $3.89~{\rm  TeV},-4.10~{\rm TeV},
 19.4, 0.039)$ for $(i, j) = (L, R)$ and $(R, L)$, respectively, corresponding to the maximum entropy which suggest the SUSY breaking scale to be about $5$ TeV, thereby allowing considerable low values of sparticles masses than the flavour-conserving CMSSM.
\end{abstract}
\end{center}
\end{titlepage}

\section{Introduction}
 
The recent discovery of Higgs-Boson at the Large Hadron Collider (LHC)~\cite{Aad:2015zhl} is quite crucial as it provides mass to the constituent particles of the Standard Model (SM)~\cite{Djouadi:2005gi}, which has been a successful theory in experiments. However, the SM lacks a stability mechanism that could protect the Higgs mass from overgrowing against the radiative corrections. Besides, it also lacks explanations for some of the interesting phenomena of nature, such as the existence of dark matter, baryogenesis, neutrino oscillations, and grand-unification, which give hints for extending SM. Among the possible extensions of the SM, supersymmetry (SUSY)~\cite{Martin:1997ns, Tata:1997uf, Drees:1996ca, Aitchison:2005cf, Fayet:2015sra, 
Djouadi:2005, Cane:2019ac, Allanchach:2019wrx} is still considered one of the most important candidate theories due to its elegance in addressing some of the aforementioned problems. The minimal supersymmetric Standard Model (MSSM)~\cite{Aitchison:2005cf, Djouadi:2005, Tanabashi:2018oca, Dawson:1997tz, Heinemeyer:1998np, Draper:2016pys} involves the Higgs sector containing five Higgses, namely two CP-even and one CP-odd Higgs-Bosons ($h$, $H$, and $A_0$) and two charged Higgs-Bosons $(H^{\pm})$. The inclusion of non-minimal flavour violation (NMFV) in the squark-sector of MSSM allows significant contributions to the mass of Higgs-Bosons through additional flavour-violating Feynman diagrams, in particular when the top-quark mixes with other flavours~\cite{AranaCatania:2011ak,AranaCatania:2012sn,Arana-Catania:2014ooa}. These additional contributions could grow as high as 5 to 10 GeV or so depending upon the value of flavour-violating couplings of the top-quark. It is therefore expected that the SUSY breaking scale reduces significantly compared to MSSM without such flavour-violating interactions while yielding mass of the lighter CP-even Higgs-Boson to its observed value at the LHC~\cite{Aad:2015zhl}.

Non-minimal flavour-violating extension of MSSM in the squark sector has already been studied in detail~\cite{Heinemeyer:2004by,AranaCatania:2011ak,AranaCatania:2012sn,Arana-Catania:2014ooa,DeCausmaecker:2015yca,Bernigaud:2018qky,Hu:2019heu,Hahn:2005qi, Carena:2006ai, Bruhnke:2010rh, Herrmann:2011xe, Fuks:2011dg, Fuks:2008ab, delAguila:2008iz, Bernigaud:2018vmh,Bozzi:2007me,Guasch:1999jp, Cao:2006xb, Cao:2007dk,Hiller:2008sv,Kowalska:2014opa}. 
For example, Ref.~\cite{Heinemeyer:2004by} discusses the effects of up-type squarks under the second and third generations and its consequences on the electroweak precision 
observables (EWPOs) and the mass of lighter MSSM Higgs-Boson. The effect of such interactions has further been examined for the processes of flavour-changing neutral-current (FCNC)~\cite{Chung:2003fi} analogous to NMFV and in the context of LHC signatures~\cite{Bozzi:2007me, DeCausmaecker:2015yca, Guasch:1999jp, Cao:2006xb, Cao:2007dk}. 
Ref.~\cite{Bernigaud:2018qky} demonstrates the link between the GUT and the TeV scale for flavour-violating terms in the squark and the slepton sectors. 
The flavour-violating interactions have also been studied for the other MSSM extensions, such as for GMSB model~\cite{Fuks:2008ab}, AMSB model~\cite{Fuks:2011dg}, $Z_3$ invariant NMSSM~\cite{Hu:2019heu},  
hybrid gauge-gravity model~\cite{Hiller:2008sv}, and phenomenological MSSM model~\cite{Kowalska:2014opa}. Moreover, the role of supersymmetric particles in the  flavour-changing decays of the MSSM neutral Higgs-Bosons in favour of second and third generation quarks has been discussed in Refs.~\cite{Curiel:2002pf,Curiel:2003uk,Bejar:2004rz}.

The main focus of the current study involves analysing the consequences of such flavour-violating terms for the constrained MSSM (CMSSM)~\cite{Kane:1993td,allenach,Balazs:2013qva,Fowlie:2012im,Athron:2017fxj,GAMBIT:2017snp,Ellis:2018jyl} framework using information entropy of the lighter CP-even Higgs-Boson as a tool with the assumption that it corresponds to the observed Higgs-Boson at the LHC.

Information entropy has been effectively implemented to explore particle physics and has achieved notable outcomes. For example, in Ref.~\cite{dEnterria:2012eip}, the accurate assessment of SM Higgs mass by maximising the product of its branching ratios has well consented to the experimental value observed at the LHC. Moreover, this predicts that there is a physical phenomenon related to assessing the significant mass of the Higgs-Boson distributed at its maximum possible decays. Correspondingly, in Ref.~\cite{Alves:2014ksa}, the Higgs mass is constructed through the maximum entropy principle (MEP) and the information entropy is evaluated by taking branching fractions of the SM Higgs-Boson. MEP has been satisfactorily implemented in a variety of contexts, such as the exploration of new modes of decay of the Higgs-Boson at the LHC~\cite{Alves:2020cmr}, axion mass assessment in axion-neutrino interaction using effective field theory models~\cite{Alves:2017ljt}, and analysing of the CMSSM~\cite{Gupta:2020whs}. Furthermore, the work of Refs.~\cite{Millan:2018fme,Llanes-Estrada:2017clj} has shown that Gibbs-Shannon entropy is constituted for devising distributions through the decays of hadrons and assessing the information relating to the new decay channel added to it.

The paper is organised as follows. In Section II, we discuss the flavour-violating MSSM and its consequences for the Higgs-Boson and other sparticles. In section III, we discuss the role of information entropy of the Higgs-Boson in the context of the model under consideration. A detailed analysis of the flavour-violating CMSSM using information theory is presented in section IV. Finally, we summarise our findings in Section V.

\section{Flavour violation in MSSM}
Allowing flavour-violation interaction in the squark sector would impact the CMSSM through the enriched squark mass matrix and the trilinear coupling matrix with additional flavour-violating off-diagonal terms, which could be given in terms of the up-type trilinear coupling as 

\beq\label{eq:y:1}   
v_u {\cal A}^u=
\begin{pmatrix}
	0& 0 & 0 \\
	0 & 0 &\delta^{LR}_{ct}M_{\tilde U_{L22}}M_{\tilde U_{R33}}\\
	0 &  \delta^{RL}_{ct}M_{\tilde U_{R22}} M_{\tilde U_{L33}}& m_t A_t\\
\end{pmatrix},
\quad
\eeq   
where the dimensionless parameter, $\delta^{ij}_{XY}$, is the flavour-violating coupling, ${X}$ and ${Y}$ represent the generations of squarks while $i$, $j$ represent $\emph{L}$ or $\emph{R}$ chirality. The $\delta^{RL}_{ct}$ and $\delta^{LR}_{ct}$ are the flavour-violating couplings of the top-quark with the charm-quark, $v_u$ is the vacuum expectation value of the $H_u$ field, ${\cal A}^u$ matrix corresponds to the trilinear-type couplings between the Higgs and the up-type squarks, $m_t$ is the top-quark mass, $A_t$ is the stop trilinear coupling parameter, and $M_{\tilde U_{LXX}}$ or $M_{\tilde U_{RYY}}$ denotes as up-type squark mass. The off-diagonal terms containing the couplings of $\emph{LR}$ and $\emph{RL}$ of the top-charm sector give the largest contributions to the Higgs mass due to the direct Yukawa type coupling between Higgs-Bosons and scharm-stop quarks. 
It is to be noted that in the current study, we explore CMSSM while only the aforementioned flavour-violating couplings are present according to Refs.~\cite{AranaCatania:2011ak,AranaCatania:2012sn,Arana-Catania:2014ooa} as the effects of flavour-violating couplings of top-quark with an up-quark are suppressed by a factor of about $ \frac{m_u}{m_c}$ ($m_u$ and $m_c$ representing masses of up-quark and charm-quark, respectively) as compared to the couplings of top-quark with a charm-quark. The inclusion of flavour-violating terms modify the mass-matrices of left-chiral up-type squarks to 
 
 \beq\label{eq:y:2}   
{\cal M}^2_{\tilde U_L} =
 \begin{pmatrix}
 	M^2_{\tilde U_{L{11}}}& 0 & 0 \\
 	0 & M^2_{\tilde U_{L{22}}} & \delta^{LL}_{23}M_{\tilde U_{L{22}}}M_{\tilde U_{L{33}}}\\
 	0 &  \delta^{LL}_{23}M_{\tilde U_{L22}}M_{\tilde U_{L{33}}}& M^2_{\tilde U_{L{33}}} \\
 \end{pmatrix}
 \quad 
 \eeq
and similarly for the right-chiral up-type squarks $ {\cal M}^2_{\tilde U_R}$. In the above equation, $\delta^{LL}_{23}$ represents the flavour-violating coupling of the top-quark with the charm-quark in the LL sector. These above-mentioned terms are present at the electroweak scale down the SUSY breaking scale with the help of renormalisation group equations (RGEs).
Moreover, in this NMFV scenario, the procedure is to take the off-diagonal flavour-violating terms of the squark mass matrices and the trilinear coupling matrices having the Super-CKM basis. Similarly, the down-type mass-matrices ${\cal M}^2_{\tilde D_i}$ and down-type trilinear coupling matrix $v_d {\cal A}^d$, where $v_d$ is the vacuum expectation value of the $H_d$ field and ${\cal A}^d$ matrix corresponds to the trilinear-type couplings between the Higgs and the down-type squarks, can be build through the above matrices only by interchanging the respective indices.

With the addition of flavour-violating coupling to the CMSSM, one-loop radiative correction to the Higgs-Boson mass is as follows
\beq 
\label{eq:y:3}
\triangle m_{h}(\delta^{ij}_{ct})\equiv m^{NMFV}_{h}(\delta^{ij}_{ct})-m^{CMSSM}_{h},
\eeq
where, $\triangle m_{h}(\delta^{ij}_{ct})$, the value of contribution through the flavour-violating coupling, comes out to be zero when $m^{NMFV}_{h}(\delta^{ij}_{ct})$= $m^{CMSSM}_{h}$ at $\delta^{ij}_{ct}$ = 0. Both $m^{NMFV}_{h}(\delta^{ij}_{ct})$ and $m^{CMSSM}_{h}$ parameters have been evaluated through {\tt FeynHiggs 2.14.3}~\cite{Bahl:2018qog,Bahl:2017aev, Bahl:2016brp, Hahn:2013ria, Frank:2006yh, Degrassi:2002fi, Heinemeyer:1998np, Heinemeyer:1998yj} and are the Higgs masses at $\delta^{ij}_{ct}$ equal to non-zero and zero, respectively. Moreover, we only consider the impact of $\delta^{LR}_{ct}$ and $\delta^{RL}_{ct}$ as these are expected to give relatively higher (positive) contributions to the Higgs mass~\cite{AranaCatania:2011ak,AranaCatania:2012sn,Arana-Catania:2014ooa}, which could therefore reduce the SUSY breaking scale significantly  as compared to $\delta^{LL}_{23}$ and $\delta^{RR}_{23}$. 

\section{Information Entropy of the Higgs-Boson}
Shannon~\cite{shannon,jaynes:1957,thomas:2006} investigated entropy as a measure of uncertainty with respect to the information content. The information entropy (or Shannon's entropy) can be given by Eq. 2 of Ref.~\cite{Gupta:2020whs} and is well explained in the paper, along with a thorough understanding of the MEP in the context of Higgs-Boson in the CMSSM. The information theory is a theory of probability in which each probability describes the uncertainty of an event. Thus, the probability distribution of the system depicts the information about each event. Each probability can vary from zero to one, non-zero values of probability give the uncertainty. Also, these events are mutually independent and exhaustive. Probable events provide less information as compared to rarer events that give more information, while the result of an event is previously known so it provides no new information or zero entropy. 
Maximum entropy signifies a state of equilibrium and the system's maximum uncertainty. The MEP is capable of determining the best estimate of the variable associated with the probability distribution deduced from Shannon's entropy of the system. For our study, we evaluate the Higgs entropy concerning the branching ratios of the Higgs decays using Shannon's entropy. The measurement of Higgs mass by maximising Higgs entropy corresponds well to the mass observed at the LHC~\cite{Aad:2015zhl}. Further, the Higgs entropy is used to determine the mass of sparticles effectively. For this, we consider an ensemble containing $\cal N$-number of independent Higgs-Bosons detected at the LHC. Here, each of the Higgs-Bosons is feasible for decay into available decay modes with probabilities $p_{_q} (m_h)$ of the branching ratio $Br_q (m_h)$ contributed to each decay channel which can be expressed as 
 \beq
\label{eq:y:4} 
 p_{_q} (m_h) \equiv Br_q (m_h) =  \frac{\Gamma_q (m_h)}{\Gamma_h (m_h)}, 
 \eeq
where the partial decay width of the Higgs-Boson to $q^{th}$ decay mode is indicated as $\Gamma_q (m_h)$ and considering the total decay width of the Higgs-Boson as $\Gamma_h (m_h) = \sum_{q = 1}^{n_q}\Gamma_{q}(m_h)$ while the total number of possible decay modes of the Higgs-Boson is as $n_q$.\\
The probability of an ensemble attains an ultimate state by considering its allowed decay modes in the form of multinomial distribution as follows~\cite{Alves:2014ksa}
\beq
\label{eq:y:5}
{\cal P}_{\{m{_q}\}}(m_h) = \frac{{\cal N}!}{m_1!...m_{n_q}!}\prod_{q = 1}^{n_q}{(p_{_q} (m_h))}^{m_q},
\eeq
here $\sum_{q = 1}^{n_q}{Br}_{q} = 1 $, $ \sum_{q = 1}^{n_q}m_{q} = {\cal N}$, and 
the number of Higgs-Bosons decay in the $q^{th}$ detection mode represent as $m_{q}$. The total information entropy accompanying the $\cal N$-Higgs-Bosons reaching its ultimate state using Shannon's entropy, i.e. Eq. 2 of Ref.~\cite{Gupta:2020whs}, also discussed in~\cite{Alves:2014ksa}, can thus be built by
\beq
\label{eq:y:6} 
S (m_h) = - \sum^{\cal N}_{\lbrace m_q \rbrace} {\cal P}_{\{m{_q}\}}(m_h) \ln {\cal P}_{\{m{_q}\}}(m_h).
\eeq
Consequently, an asymptotic expansion of the above-stated equation follows as the subsequent Higgs-Boson entropy, as described in~\cite{Alves:2014ksa}, can be written as
\beq
\label{eq:y:7}
S (m_h) \simeq \frac{1}{2}\ln\left(\left(2\pi {\cal N} e\right)^{n_q -  1} \prod_{q = 1}^{n_q}{p_{_q} (m_h)}\right) + \frac{1}{12 {\cal N}}\left( 3 n_q - 2 - \sum_{q = 1}^{n_q}{(p_{_q} (m_h))}^{-1}\right)+ {\cal O}\left({\cal N}^{-2}\right).
\eeq
 
\section{Flavour violation in the CMSSM and the Information Entropy}

A random sampling approach is implemented throughout the parameter space that constitutes the CMSSM in conjunction with the NMFV. The entire supersymmetric spectrum can be exemplified through the five free parameters at the GUT-scale and free flavour-violating couplings as $m_0$ unified scalar mass, $m_{1/2}$ unified gaugino mass, $A_0$ common trilinear coupling, $tan\beta$ ratio of up- and down-type Higgs-Boson vacuum expectation values (VEVs), sgn($\mu$) sign associated with the Higgsino mass parameter $\mu$, and flavour violating couplings i.e. $\delta^{LR}_{ct}$ and $\delta^{RL}_{ct}$. It is to be noted that the flavour-violating couplings $\delta^{ij}_{ct}$ are introduced at the SUSY breaking scale in the soft matrices discussed in Eqs~\ref{eq:y:1} and \ref{eq:y:2}. We use random scan over the CMSSM parameter space along with the flavour-violating couplings $\delta^{ij}_{ct}$ for the following range,
\begin{itemize}
\item $m_0 \in [0.1,6] $ TeV,
\item $m_{1/2} \in [0.1,6] $ TeV,
\item $A_{0} \in [-6,6]$ TeV,
\item $tan\beta \in [2,60]$,
\item sgn$(\mu)$ = +1,
\item $\delta^{LR}_{ct}$ or $\delta^{RL}_{ct}$  $\in$ [$-$0.07,0.07].
\end{itemize}

Utilising these above parameters, the supersymmetric spectrum has been generated through {\tt Softsusy 4.1.3}~\cite{Allanach:2001kg} and then interfaced with {\tt FeynHiggs $2.14.3$}~\cite{Bahl:2018qog,Bahl:2017aev, Bahl:2016brp, Hahn:2013ria, Frank:2006yh, Degrassi:2002fi, Heinemeyer:1998np, Heinemeyer:1998yj} to evaluate the Higgs observables such as masses, branching ratios, and $\rho$-parameter. Using the spectrum generator, {\tt Superiso v4.0}~\cite{Arbey:2018msw} computes the muon anomalous magnetic moment and the B-physics branching ratios, and {\tt micromegas 
5.0.4}~\cite{Belanger:2004yn, Belanger:2001fz} measures the relic density of the neutralino dark matter. Concerning the information theory that opens the potential of knowing the findings using only the knowledge of the branching ratios of the Higgs-Boson and then enlightening the Higgs mass regarded as unknown. Moreover, the preferred mass of sparticles can be assessed effectively. Thus we estimate the entropy of an ensemble consisting of neutral CP-even lighter Higgs-Bosons in the NMFV framework that is driven by Eq.~\eqref{eq:y:7}. For achieving this, we need to compute the available decay modes of the CP-even lighter Higgs-Boson as $h\rightarrow\gamma\gamma$, $h\rightarrow \gamma Z$, $h\rightarrow Z Z^*$, $h\rightarrow W W^*$, $h\rightarrow gg$, $h\rightarrow f\bar{f}$ with $f\in \{u, d, c, s, b, e^\pm, \mu^\pm, \tau^\pm\}$. The aforementioned branching ratios of the Higgs-Boson are assessed with {\tt FeynHiggs 2.14.3}~\cite{Bahl:2018qog,Bahl:2017aev, Bahl:2016brp, Hahn:2013ria, Frank:2006yh, Degrassi:2002fi, Heinemeyer:1998np, Heinemeyer:1998yj}. In accordance with an information-theoretic method, the entropy is maximised for a specific mass of the Higgs-Boson while considering all its available decay modes of the Higgs-Boson. Afterwards, marginalisation on all other MSSM parameters, marginalised entropy, S, varies with $m_h$ only and then scaled by a $1/S_{max}$ normalisation factor. Our parameter space is then limited by several constraints, including LEP data on sparticles masses, i.e. neutralino and chargino masses and Higgs mass, confines from $ \bigtriangleup\rho$, $\triangle a_{\mu}$, $ BR(b \to s\gamma)$, $ BR(B^0_s \to \mu^+\mu^-)$, and relic density of dark matter $\Omega_{\chi}h^2$ at 2.5$\sigma$ confidence level with experimental values as described in Table~\ref{tab:table1}.  

\begin{table}[h]
 \begin{centering}
    \tabcolsep 0.4pt
    \footnotesize
    \begin{tabular}{cccc}
        \hline
    \textbf{Constraint}&\textbf{Observable}  & \textbf{ Experimental~Value}& \textbf{ Source}	\\
       \hline 
LEP &$ m_{h}$ &  $>$ 114.4 GeV& \cite{Schael:2006cr}\\
&$ m_{\tilde\chi^{0}_{1,2,3,4}}$ & $>$ 0.5 $m_Z$&  \cite{Zyla:2020zbs}\\
&$ m_{\tilde\chi^{\pm}_{1,2}}$& $>$ 103.5 GeV&  \cite{Zyla:2020zbs}\\
 \hline 
PO & $ \triangle\rho$  & $0.00038\pm0.0002$&  \cite{Zyla:2020zbs}\\
&$ BR(b \to s\gamma)$ & $(3.32\pm0.15)\times10^{-4}$&  \cite{Zyla:2020zbs,Amhis:2019ckw}\\
&$BR(B^0_s \to \mu^+\mu^-)$& $(3.0\pm0.4)\times10^{-9}$&  \cite{Zyla:2020zbs}\\
& $\triangle a_{\mu}$ & $(2.51\pm0.59)\times10^{-9}$&\cite{Abi:2021gix}  \\
 \hline 
DM & $ \Omega_{\chi}h^{2}$ & $0.1200\pm0.0012$& \cite{Zyla:2020zbs}\\
       \hline
    \end{tabular}
     \caption{\sf{A list of several experimental observables used as constraints.}}
      \label{tab:table1}
   \end{centering}
\end{table}

The impact of NMFV is examined in the $\emph{LR}$ and $\emph{RL}$ sectors separately of the scharm-stop flavour-violating interaction within the CMSSM model. While the CMSSM acquires the one-loop Higgs mass at 125 GeV with the high supersymmetric breaking scale, which is far from being observed at the LHC. Therefore, with the preface of NMFV in the scharm-stop flavour-violating interaction, the above difficulty is found to sort with the reduced supersymmetric scale at a 125 GeV Higgs mass. Comparatively, we report the outcomes in terms of information entropy within different sets of constraints, namely (a) LEP, (b) LEP$+$PO, and (c) LEP$+$PO$+$DM. 

\begin{figure}
\begin{centering}
\includegraphics[angle=0,width=0.48\linewidth,height=15.0em]{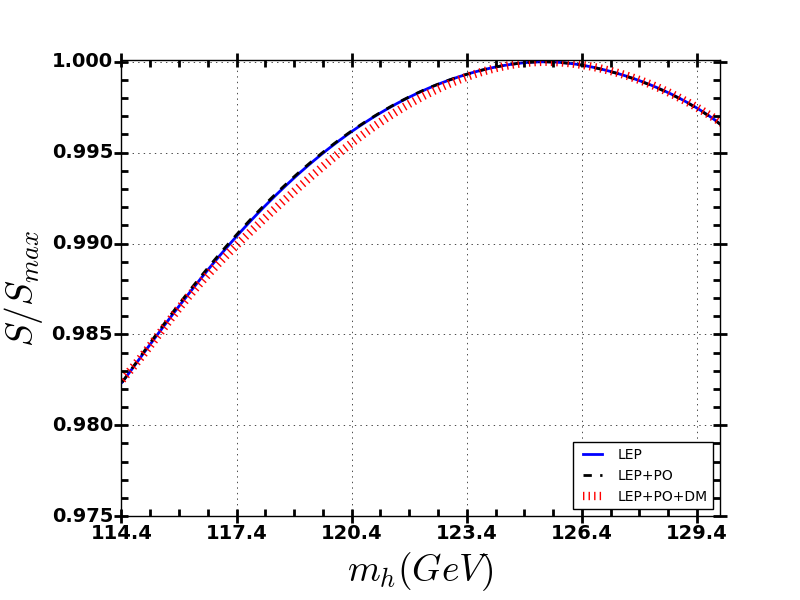}
\hspace{-2.2em}
\includegraphics[angle=0,width=0.42\linewidth,height=15.0em]{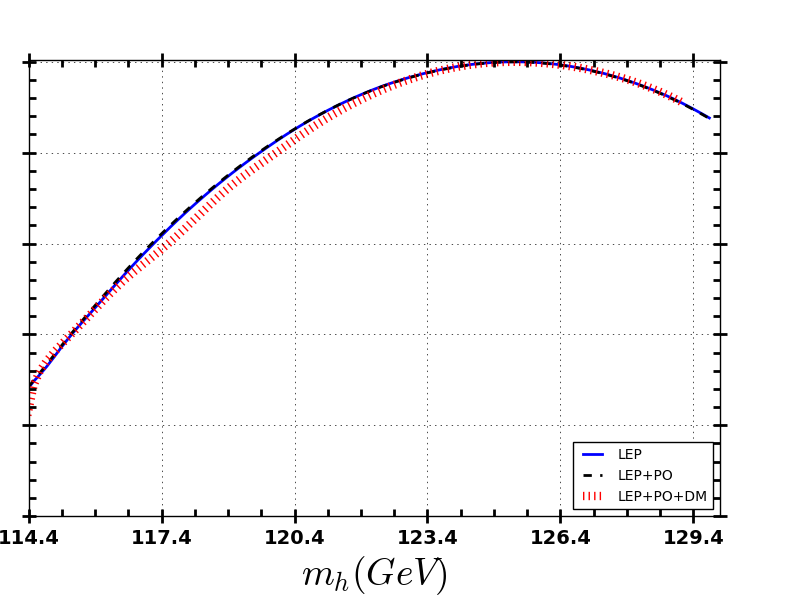}
\caption{\sf{Marginalised entropy vs lighter CP-even Higgs mass in both cases of $\emph{LR}$ (left) and $\emph{RL}$ (right) sectors. The LEP constraints are indicated by the solid blue line, the LEP and PO constraints are depicted by the black dashed line, and the LEP, PO and DM constraints are exhibited by the red dotted line. The details of the constraints are set out in Table~\ref{tab:table1}.}}
\label{fig:1}
\end{centering} 
\end{figure}

\begin{figure}
\includegraphics[angle=0,width=0.37\linewidth,height=15.0em]{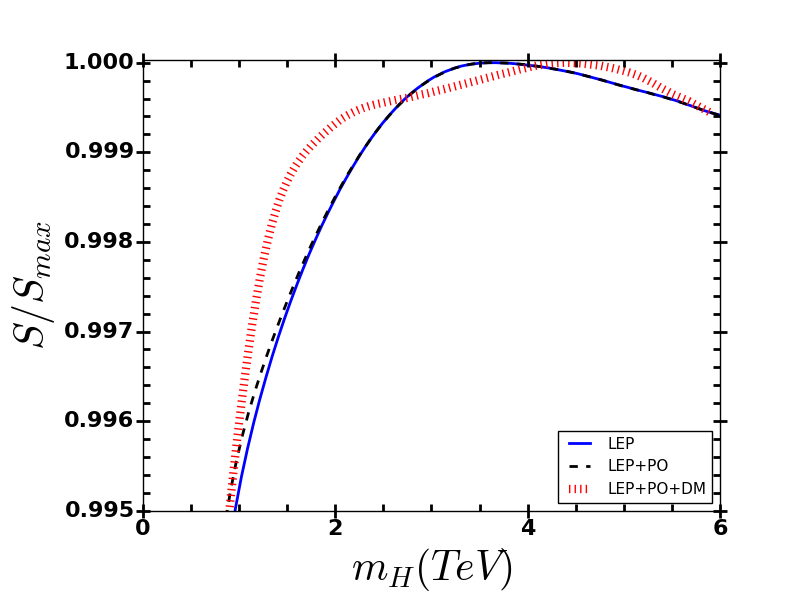}
\hspace{-1.92em}
\includegraphics[angle=0,width=0.315\linewidth,height=15.0em]{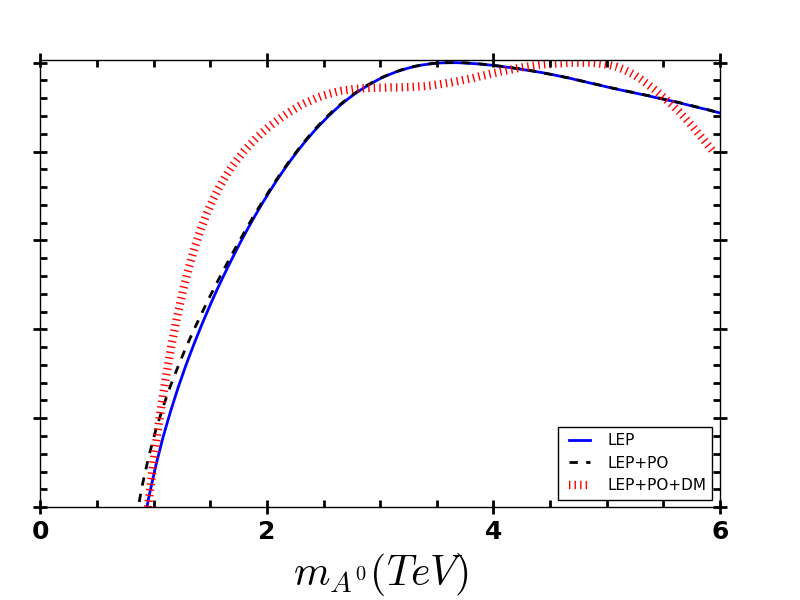}
\hspace{-1.70em}
\includegraphics[angle=0,width=0.315\linewidth,height=15.0em]{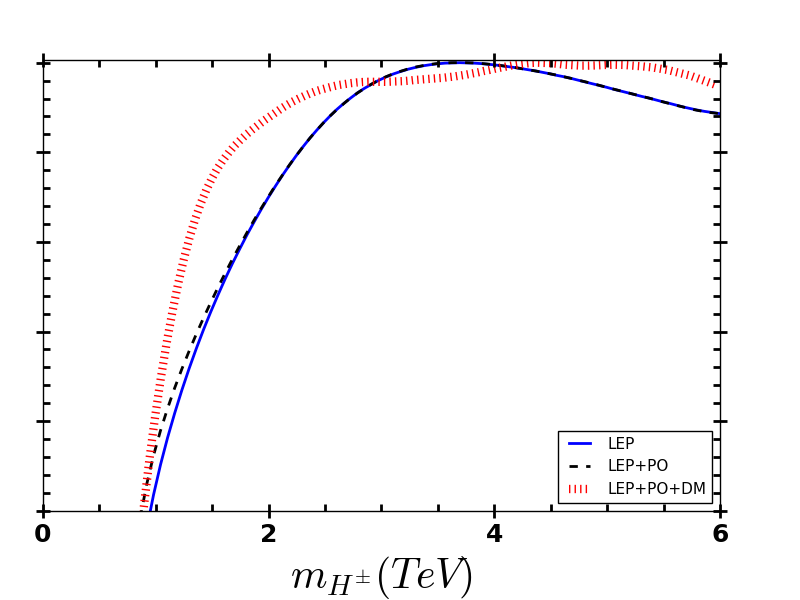}
\caption{\sf{Variation of marginalised entropy with mass of the other Higgses, i.e. (a) CP-even neutral heavier Higgs-Boson $H$ (left), (b) CP-odd neutral Higgs-Boson $A^0$ (middle), and (c) charged Higgs-Bosons $H^\pm$ (right) in case of $\emph{LR}$  sector. The colour scheme is similar to Figure~\ref{fig:1}.}}
\label{fig:2}
\end{figure}

\begin{figure}
\includegraphics[angle=0,width=0.37\linewidth,height=15.0em]{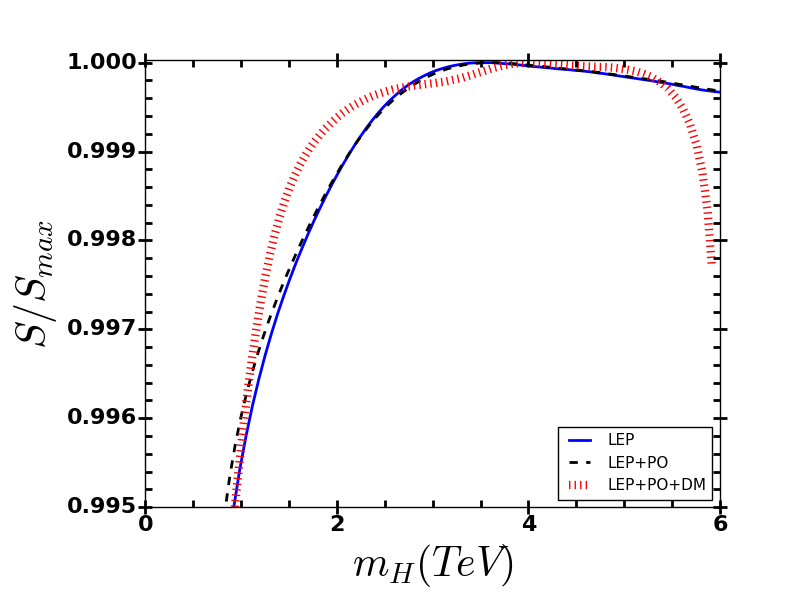}
\hspace{-1.92em}
\includegraphics[angle=0,width=0.315\linewidth,height=15.0em]{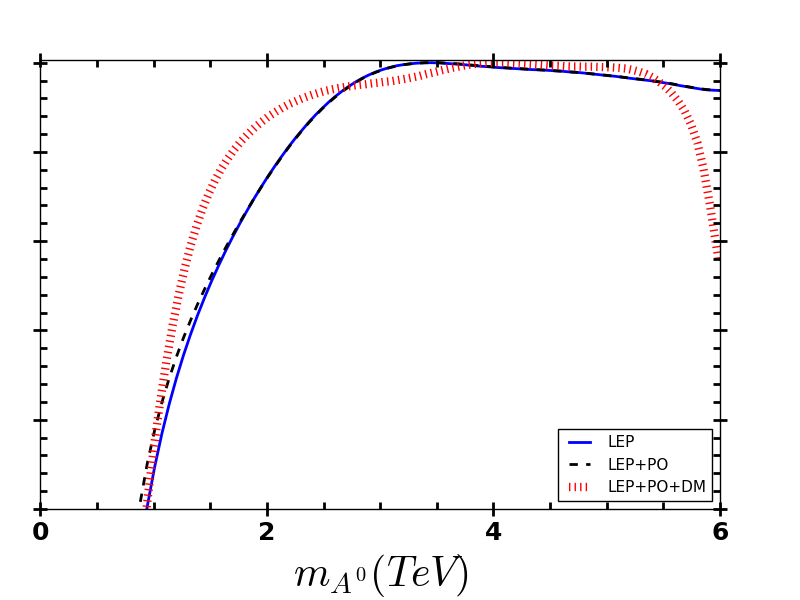}
\hspace{-1.70em}
\includegraphics[angle=0,width=0.315\linewidth,height=15.0em]{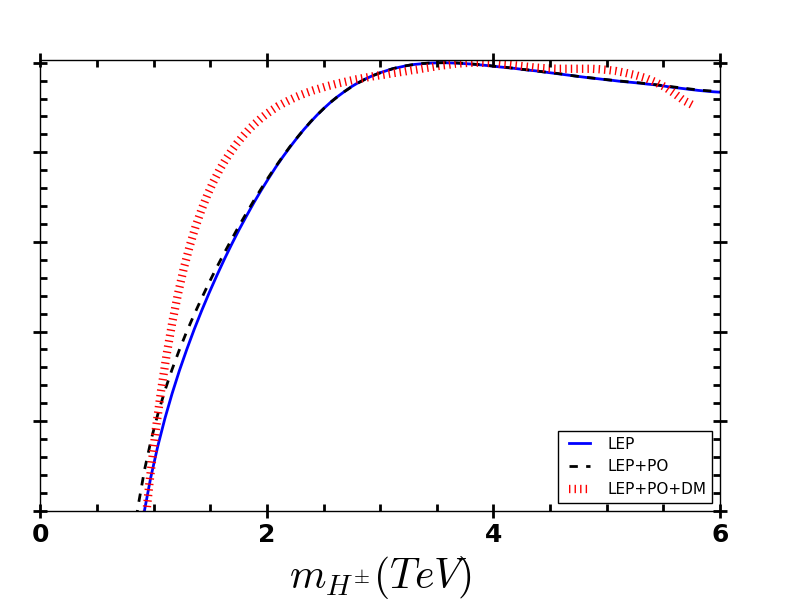}
\caption{\sf{Variation of marginalised entropy with mass of the other Higgses, i.e. (a) CP-even neutral heavier Higgs-Boson $H$ (left), (b) CP-odd neutral Higgs-Boson $A^0$ (middle), and (c) charged Higgs-Bosons $H^\pm$ (right) in case of $\emph{RL}$ sector. The colour scheme is similar to Figure~\ref{fig:1}.}}
\label{fig:3}
\end{figure}

\begin{figure}
\begin{centering}
\includegraphics[angle=0,width=0.5\linewidth,height=16.0em]{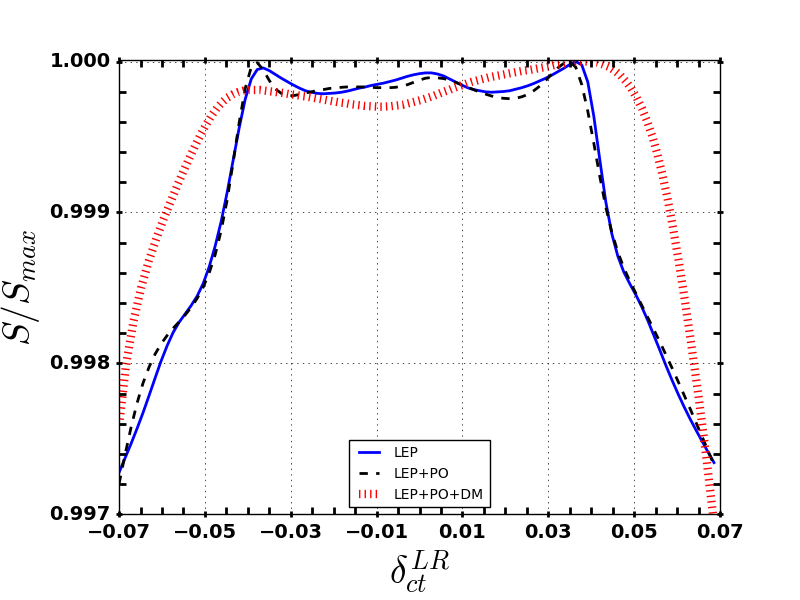}
\hspace{-2.025em}
\includegraphics[angle=0,width=0.45\linewidth,height=16.0em]{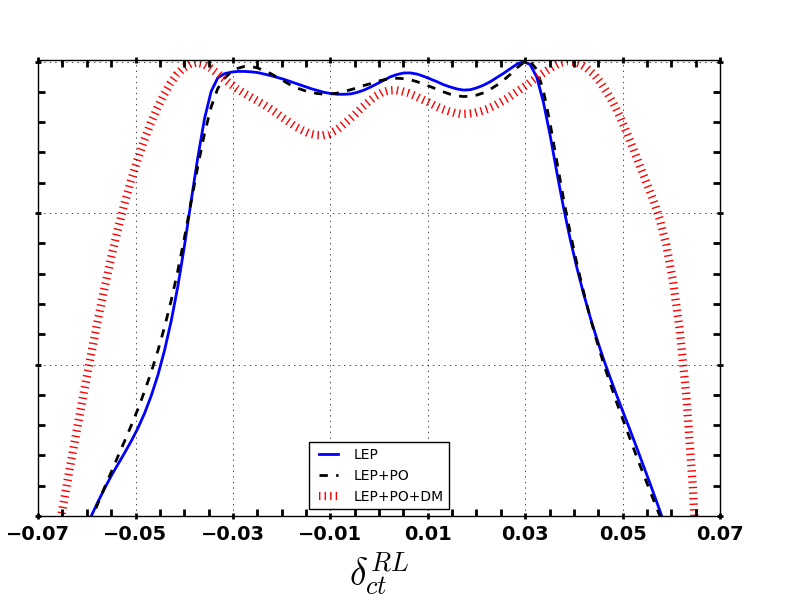}
\caption{\sf{Entropy vs flavour-violating coupling parameters, $\delta^{LR}_{ct}$ (left) and $\delta^{RL}_{ct}$ (right), for different sets of constraints. The colour scheme is similar to Figure~\ref{fig:1}.}}
\label{fig:4}
\end{centering} 
\end{figure}

\begin{figure}
\includegraphics[angle=0,width=0.39\linewidth,height=10.0em]{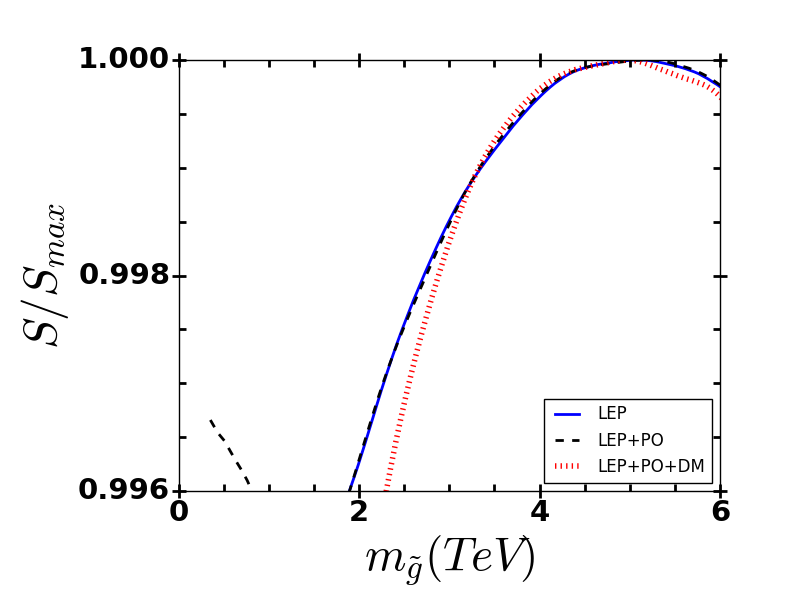}
\hspace{-4.5em}
\includegraphics[angle=0,width=0.32\linewidth,height=10.0em]{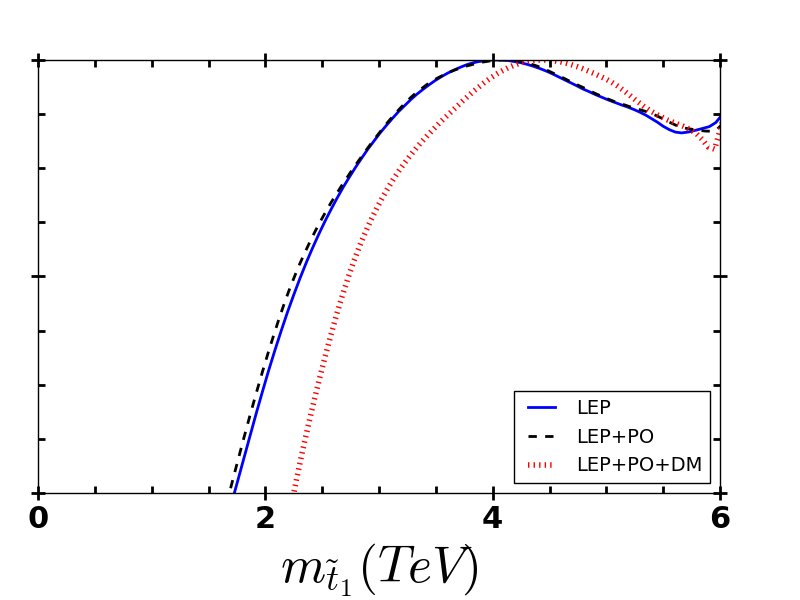}
\hspace{-4.5em}
\includegraphics[angle=0,width=0.32\linewidth,height=10.0em]{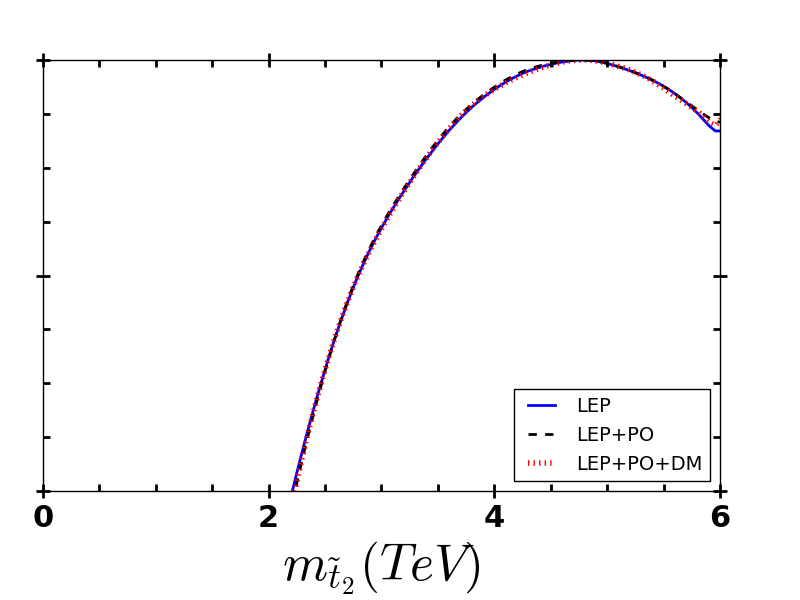}
\vspace{-0.13em}
\includegraphics[angle=0,width=0.39\linewidth,height=10.0em]{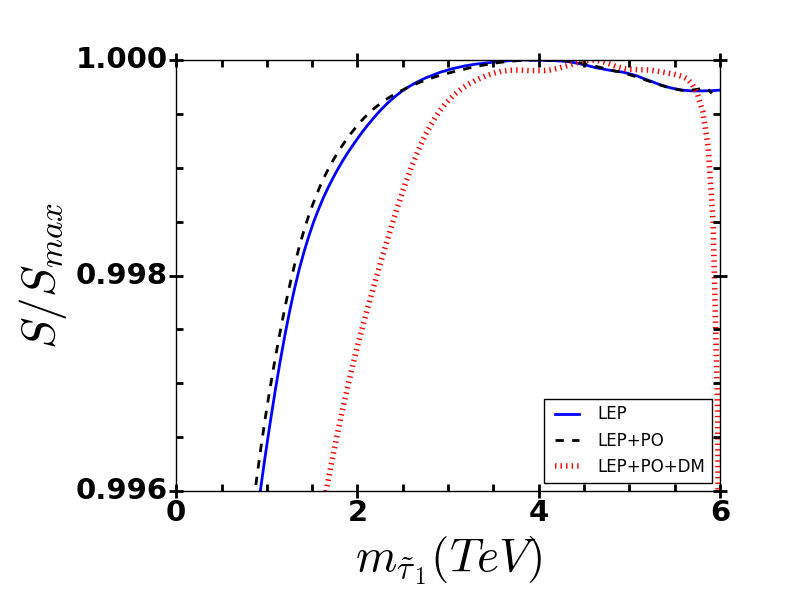}
\hspace{-4.5em}
\includegraphics[angle=0,width=0.32\linewidth,height=10.0em]{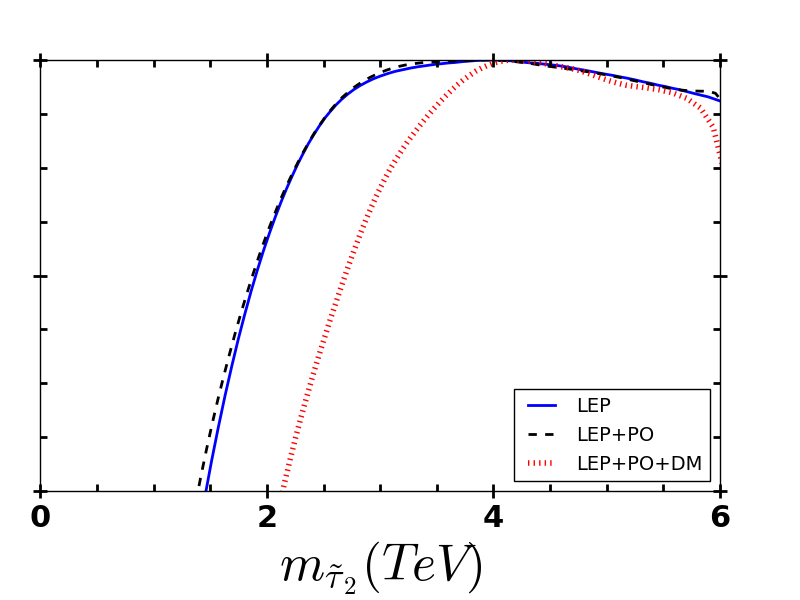}
\hspace{-4.5em}
\includegraphics[angle=0,width=0.32\linewidth,height=10.0em]{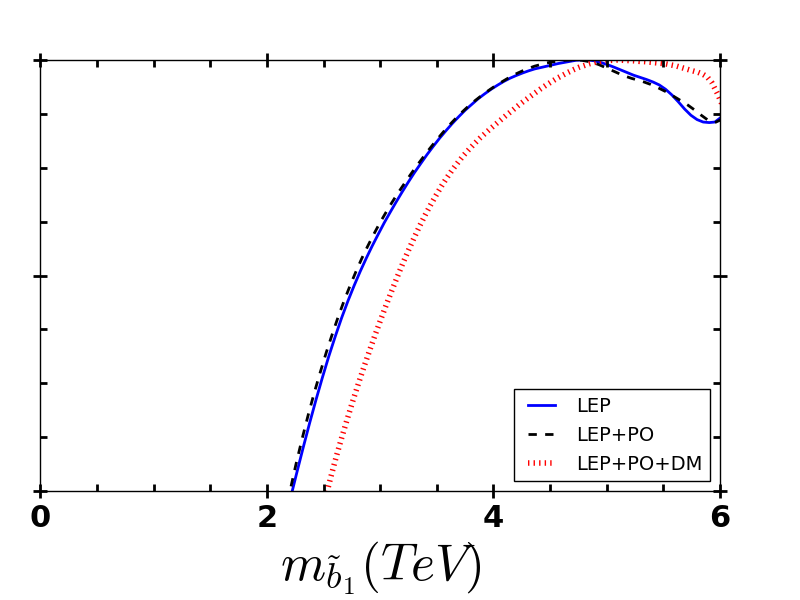}
\vspace{-0.12em}
\includegraphics[angle=0,width=0.39\linewidth,height=10.0em]{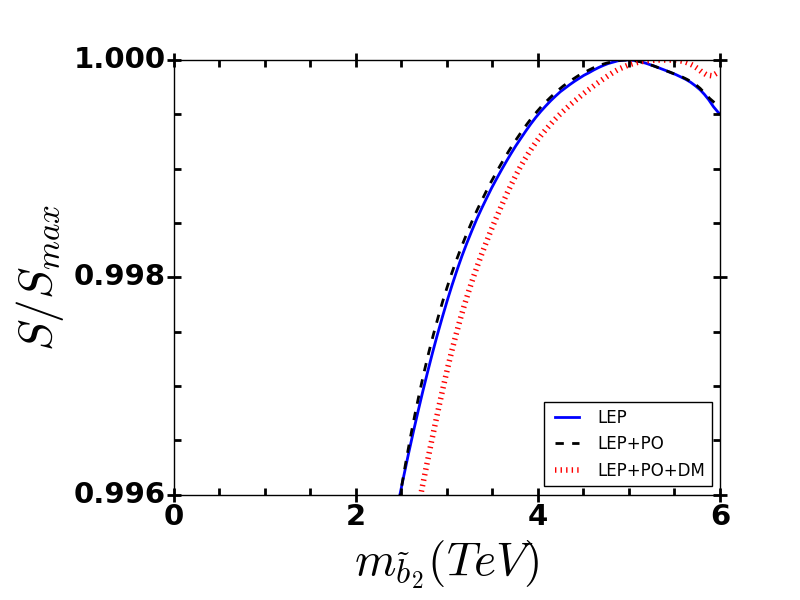}
\hspace{-4.5em}
\includegraphics[angle=0,width=0.32\linewidth,height=10.0em]{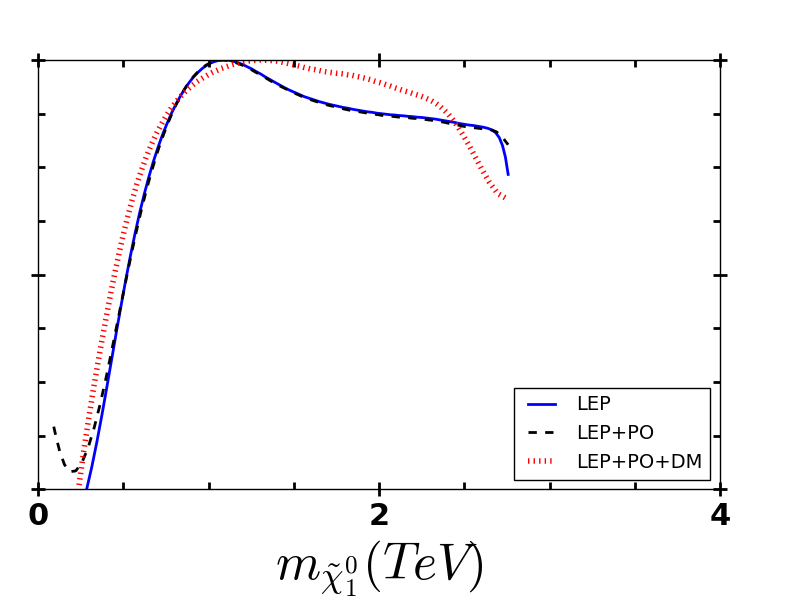}
\hspace{-4.5em}
\includegraphics[angle=0,width=0.32\linewidth,height=10.0em]{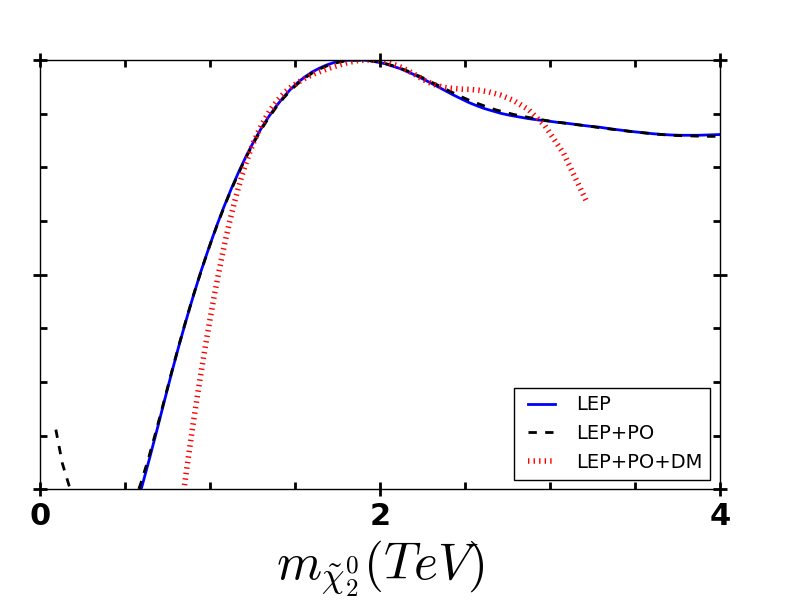}
\vspace{-0.13em}
\includegraphics[angle=0,width=0.39\linewidth,height=10.0em]{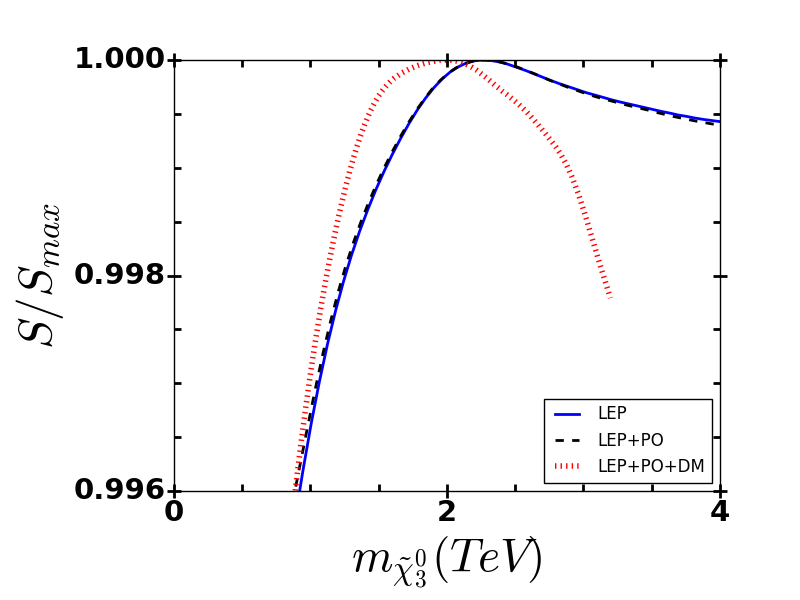}
\hspace{-1.2em}
\includegraphics[angle=0,width=0.32\linewidth,height=10.0em]{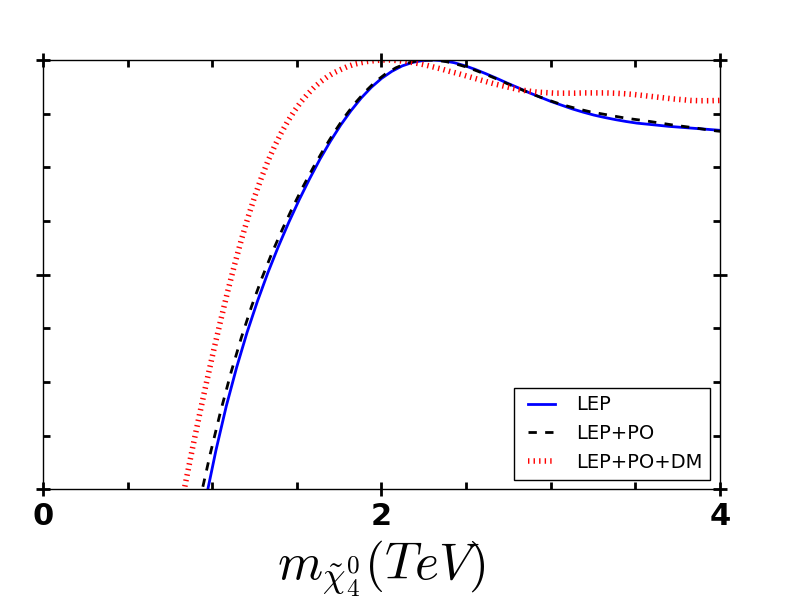}
\hspace{-1.3em}
\vspace{-0.16em}
\includegraphics[angle=0,width=0.32\linewidth,height=10.0em]{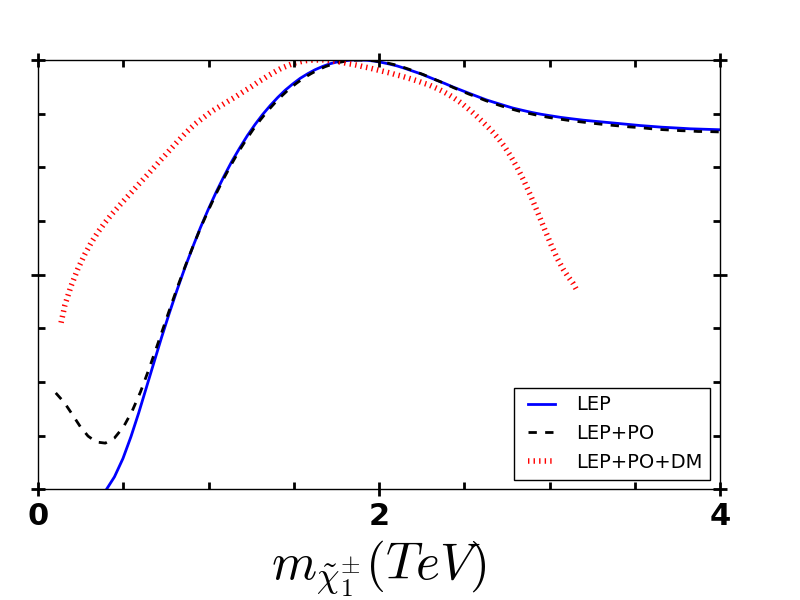}
\center
\hspace{0.38em}
\includegraphics[angle=0,width=0.40\linewidth,height=10.0em]{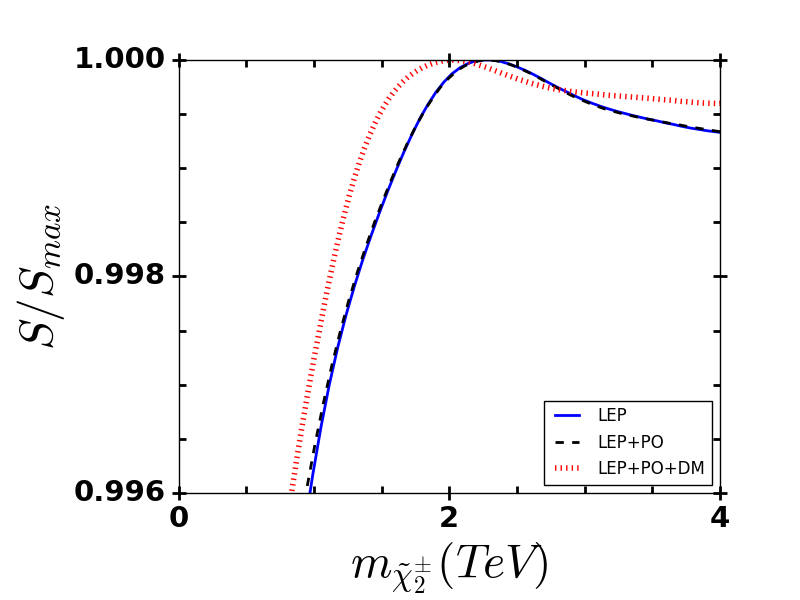}
\caption{\sf{Marginalised entropy vs sparticles mass in case of $\emph{LR}$ sector. The colour scheme is similar to Figure~\ref{fig:1}.}}
\label{fig:5}
\end{figure}

\begin{figure}
\includegraphics[angle=0,width=0.39\linewidth,height=10.0em]{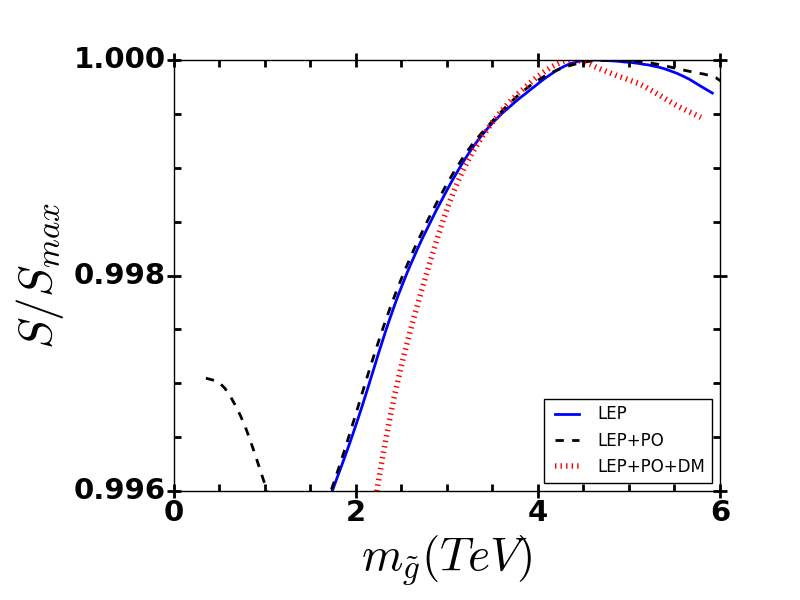}
\hspace{-4.5em}
\includegraphics[angle=0,width=0.32\linewidth,height=10.0em]{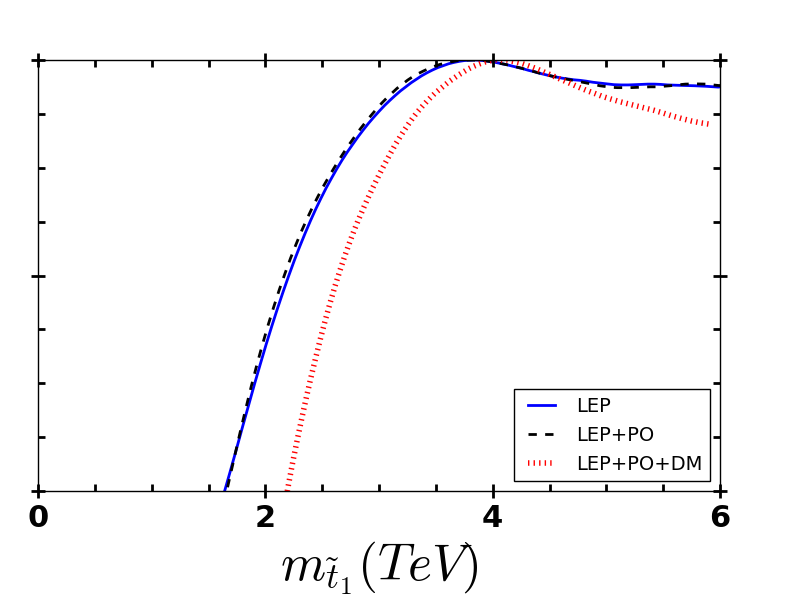}
\hspace{-4.5em}
\includegraphics[angle=0,width=0.32\linewidth,height=10.0em]{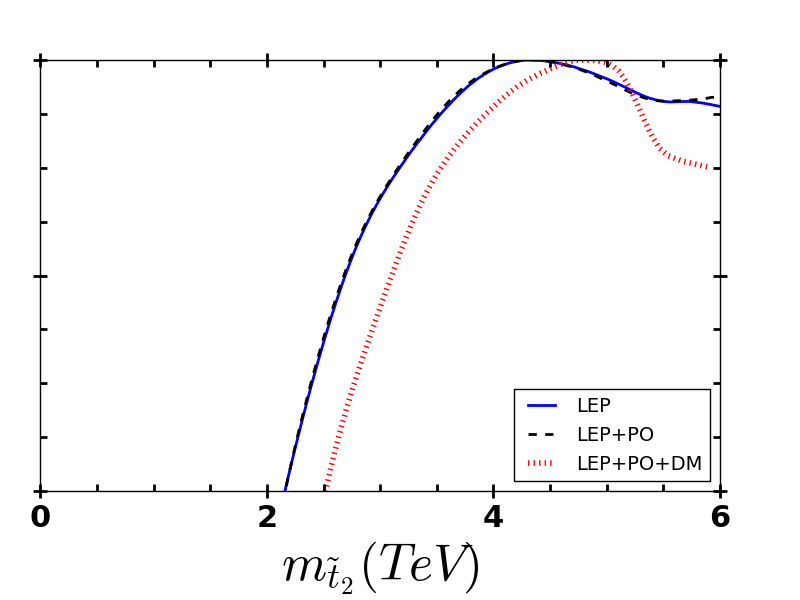}
\vspace{-0.13em}
\includegraphics[angle=0,width=0.39\linewidth,height=10.0em]{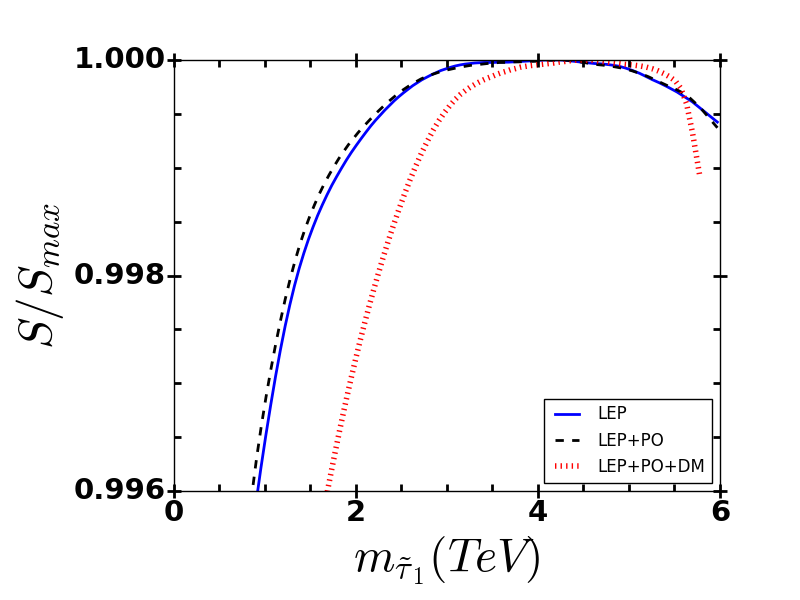}
\hspace{-4.5em}
\includegraphics[angle=0,width=0.32\linewidth,height=10.0em]{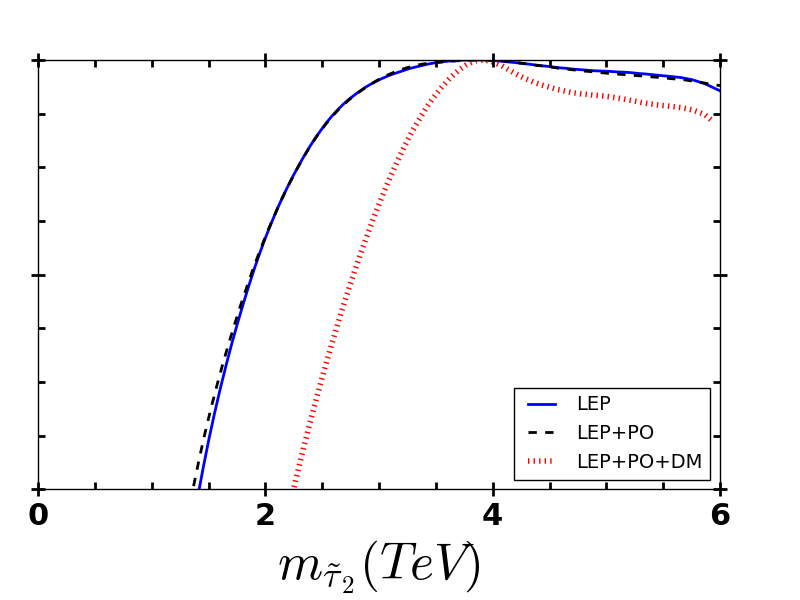}
\hspace{-4.5em}
\includegraphics[angle=0,width=0.32\linewidth,height=10.0em]{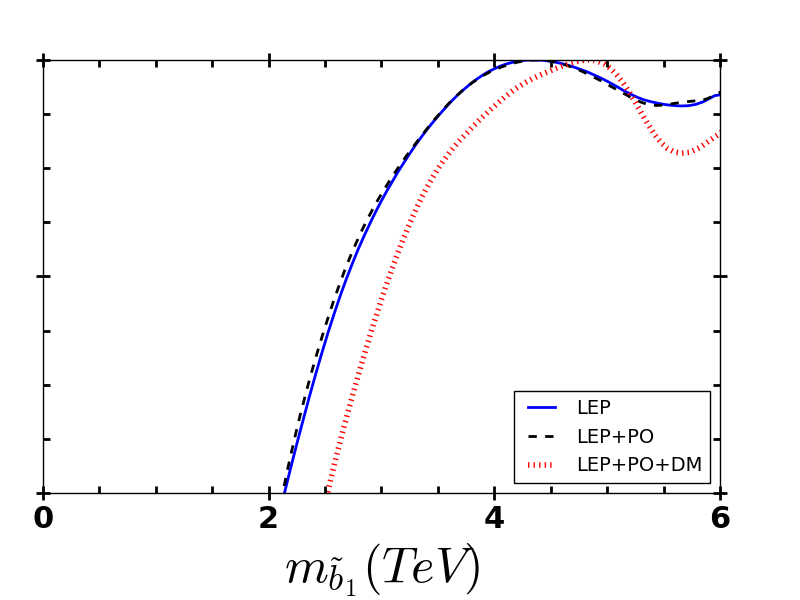}
\vspace{-0.12em}
\includegraphics[angle=0,width=0.39\linewidth,height=10.0em]{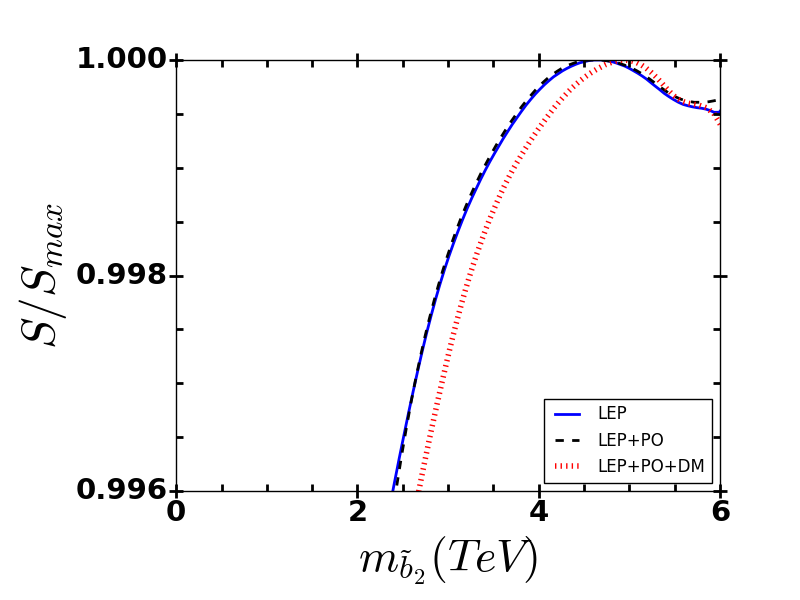}
\hspace{-4.5em}
\includegraphics[angle=0,width=0.32\linewidth,height=10.0em]{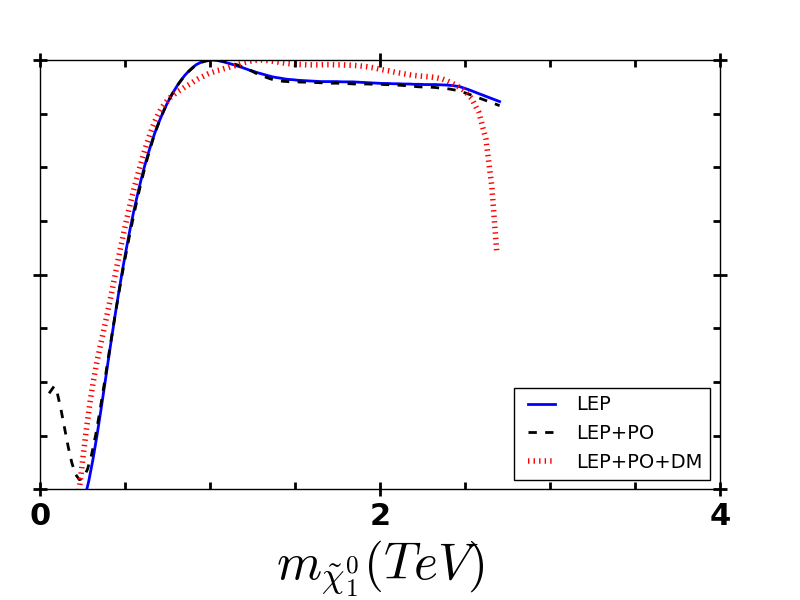}
\hspace{-4.5em}
\includegraphics[angle=0,width=0.32\linewidth,height=10.0em]{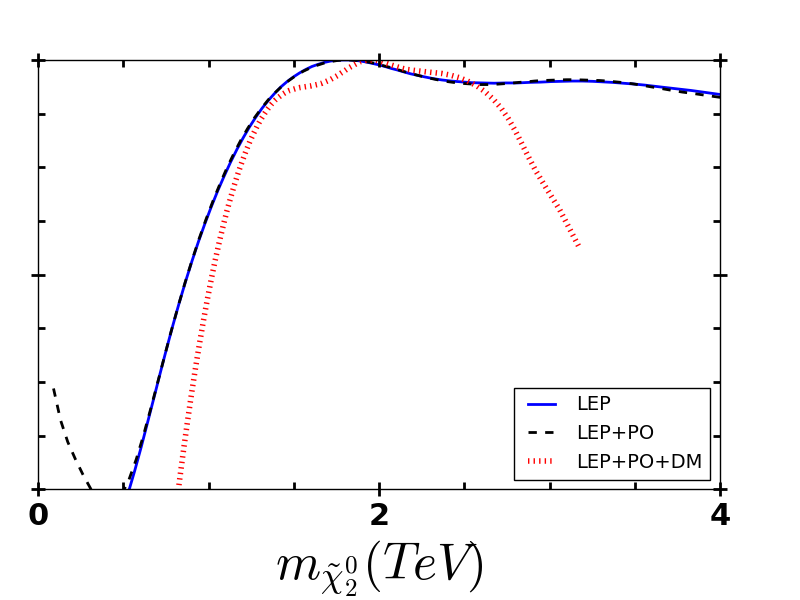}
\vspace{-0.13em}
\includegraphics[angle=0,width=0.39\linewidth,height=10.0em]{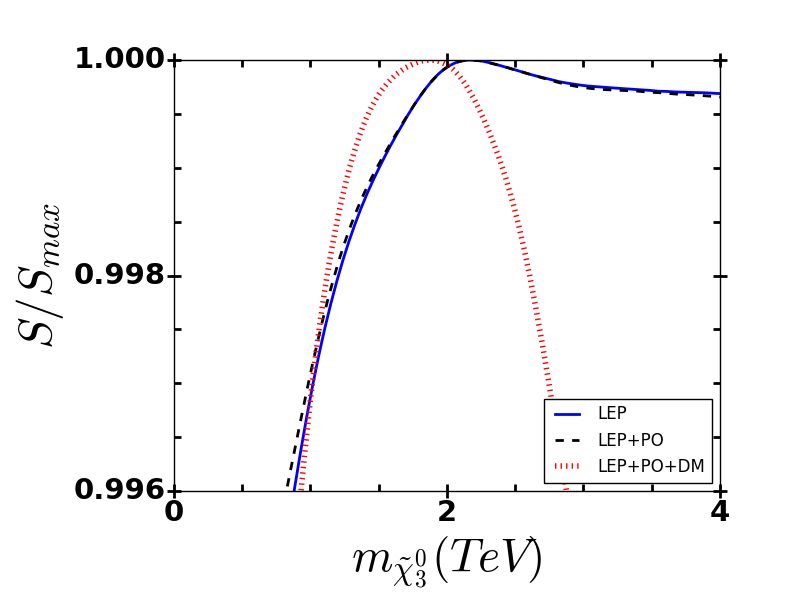}
\hspace{-1.2em}
\includegraphics[angle=0,width=0.32\linewidth,height=10.0em]{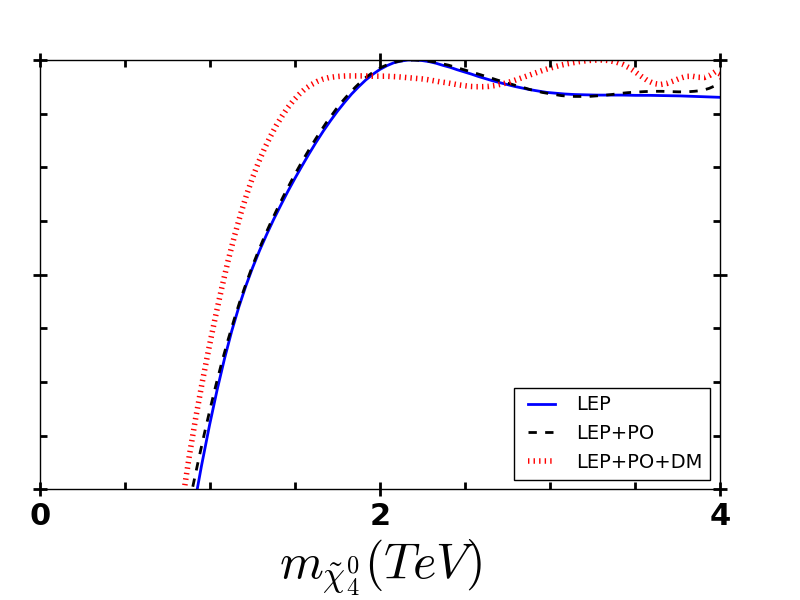}
\hspace{-1.3em}
\vspace{-0.16em}
\includegraphics[angle=0,width=0.32\linewidth,height=10.0em]{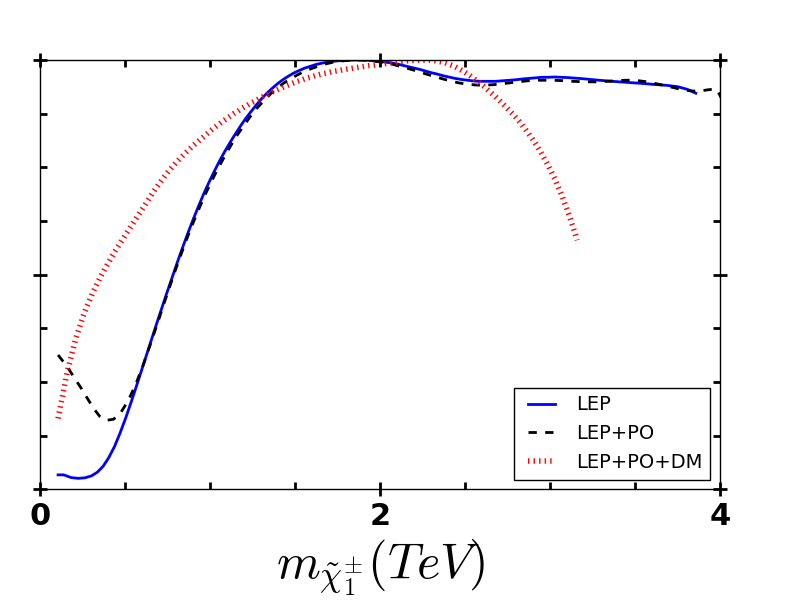}
\center
\hspace{0.38em}
\includegraphics[angle=0,width=0.40\linewidth,height=10em]{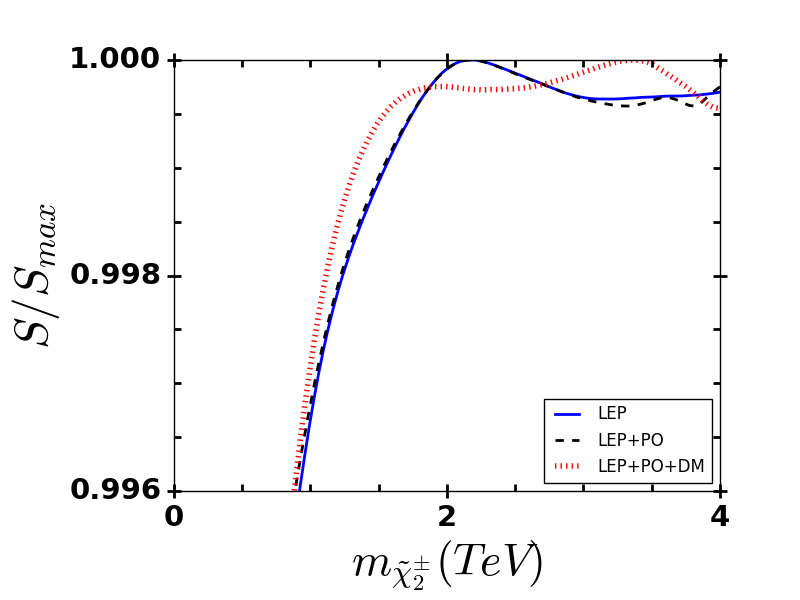}
\caption{\sf{Marginalised entropy vs sparticles mass in case of $\emph{RL}$ sector. The colour scheme is similar to Figure~\ref{fig:1}.}}
\label{fig:6}
\end{figure}

%\begin{figure}
%\vspace{-2.0em}
%\includegraphics[angle=0,width=0.55\linewidth,height=13.0em]{m0_LR.png}
%\hspace{-2.5em}
%\vspace{-0.4em}
%\includegraphics[angle=0,width=0.55\linewidth,height=13.0em]{tb_LR.png}
%\includegraphics[angle=0,width=0.55\linewidth,height=12.0em]{A0_LR.png}
%\hspace{-2.0em}
%\includegraphics[angle=0,width=0.5\linewidth,height=12.0em]{mhf_LR.png}
%\caption{\sf{Marginalised entropy vs CMSSM parameters in both cases of $\emph{LR}$ %(left) and $\emph{RL}$ (right) sectors. The colour scheme is similar to Figure~\ref{fig:1}.}}
%\label{fig:7}
%\end{figure}
\vspace{-0.3em}
\begin{figure}
\includegraphics[angle=0,width=1.0\linewidth,height=25.0em]{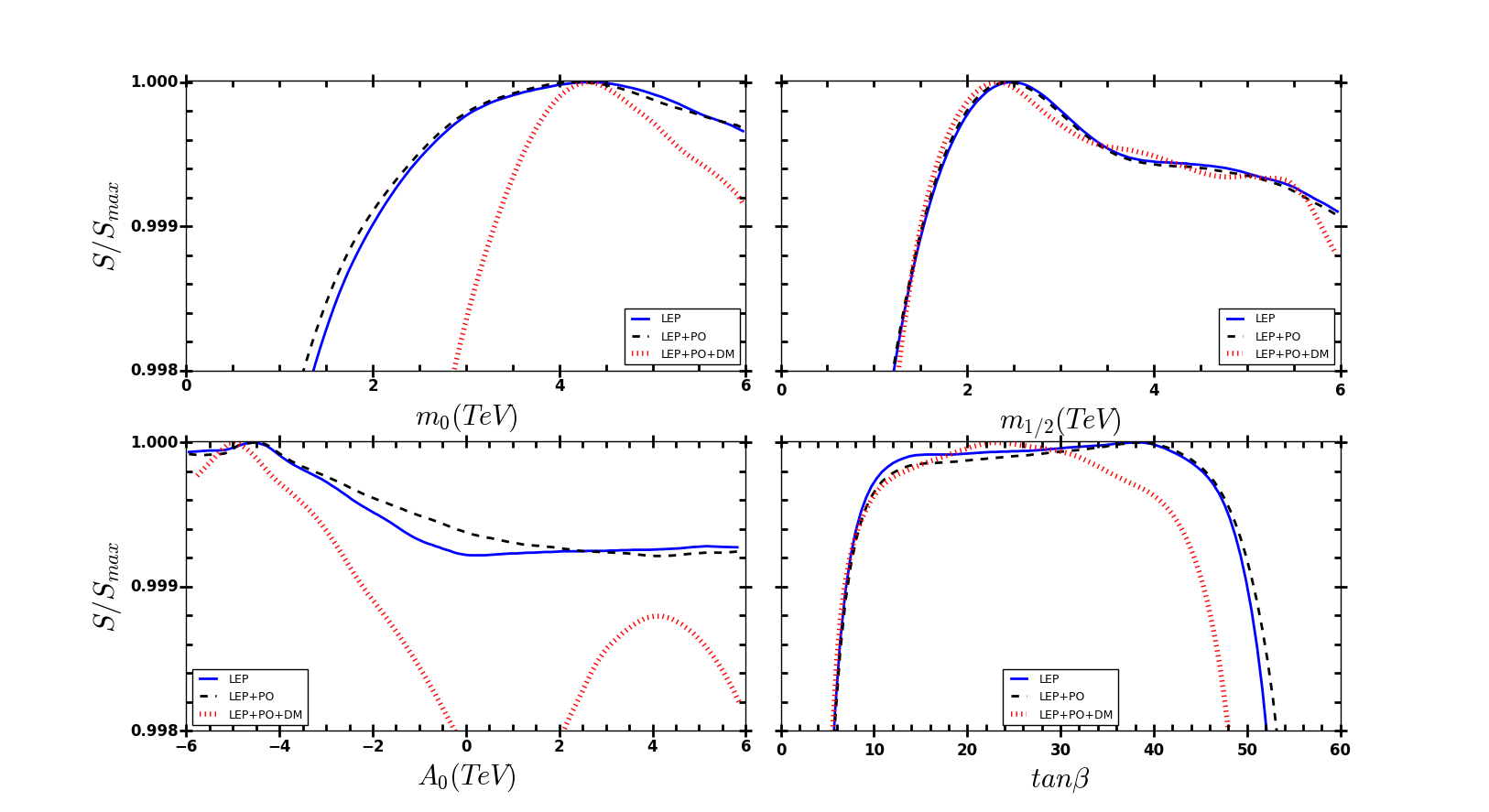}
\caption{\sf{Marginalised entropy vs CMSSM parameters in case of $\emph{LR}$ sector. The colour scheme is similar to Figure~\ref{fig:1}.}}
\label{fig:7}
\end{figure}

\begin{figure}
\includegraphics[angle=0,width=1.0\linewidth,height=25.0em]{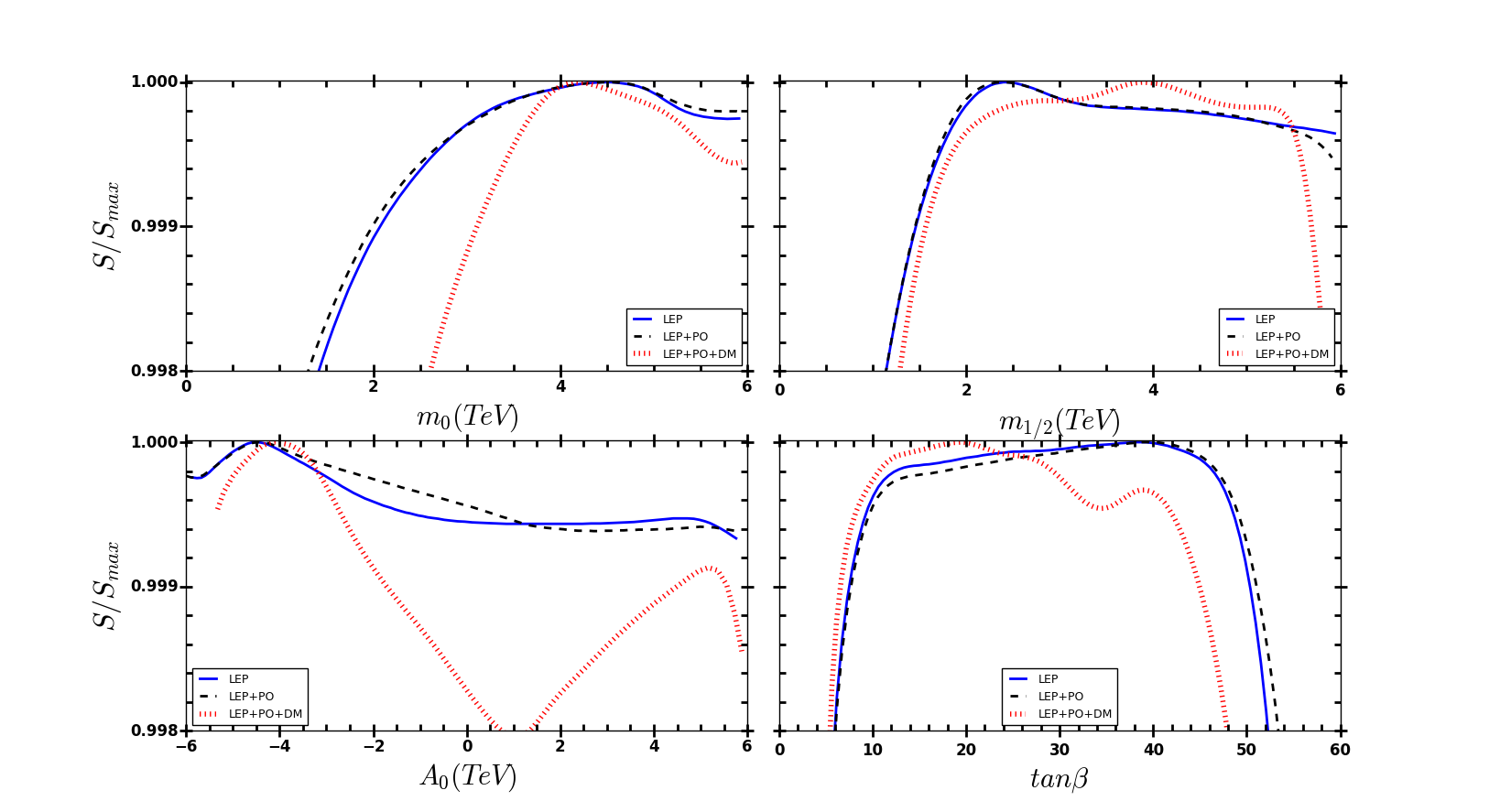}
\caption{\sf{Marginalised entropy vs CMSSM parameters in case of $\emph{RL}$ sector. The colour scheme is similar to Figure~\ref{fig:1}.}}
\label{fig:8}
\end{figure}

For deeper insight, we exhibit the variation of marginalised entropy with a CP-even lighter Higgs mass in $\emph{LR}$ (left) and $\emph{RL}$ (right) sectors, respectively, in Figure~\ref{fig:1}. Here the contribution to the Higgs mass correction is not only by $tan\beta$ and stop mass but by flavour-violating coupling also. To show the variation of other Higgses, we present the plots of marginalised entropy with the mass of the heavier CP-even neutral Higgs-Boson $H$, CP-odd neutral Higgs-Boson $A^0$, and charged Higgs-Bosons $H^\pm$ in both $\emph{LR}$ and $\emph{RL}$ sectors in Figures~\ref{fig:2} and~\ref{fig:3}, respectively. Clearly, to see the extent of NMFV in CMSSM, the marginalised entropy is plotted with scharm-stop flavour-violating interaction parameter of $\emph{LR}$ (left) and $\emph{RL}$ (right) sectors in Figure~\ref{fig:4}. Flavour-violating couplings enhance the Higgs mass by a few GeV, while the peaks of maximum entropy corresponding to these couplings are at approximately $\pm$ 0.04. These plots are symmetrical around the zero value of flavour-violating couplings. The non-zero values of flavour-violating couplings prove the impact of NMFV on CMSSM parameter space limited by LEP, PO, and DM constraints and then obtain the SUSY masses that are now reachable under LHC observation. The values of flavour-violating couplings corresponding to the peak of maximum marginalised entropy for different sets of constraints in both $\emph{LR}$ and $\emph{RL}$ sectors are outlined in Table~\ref{tab:table2}. The Higgs mass corrections may be either negative or positive based on the non-zero values of these couplings. It can even go larger than 4 GeV in negative or 1 GeV in positive correction. Consequently, there is a narrow window containing the flavour-violating couplings that contribute to the positive one-loop correction of the MSSM lightest Higgs mass. The accomplishment of the effects of NMFV can therefore be seen through Figures~\ref{fig:5} and~\ref{fig:6}, here the plots of the marginalised entropy vary with the mass of the sparticles for the sectors $\emph{LR}$ and $\emph{RL}$, respectively. The most likely values of gluino, lighter stop, lighter stau, lighter chargino, and lightest neutralino in $\emph{LR}$ sector of the scharm-stop flavour-violating interaction are 4.97 TeV, 4.39 TeV, 4.15 TeV, 1.62 TeV, and 1.30 TeV, respectively. These would be changed to 4.39 TeV, 4.02 TeV, 3.92 TeV, 2.24 TeV, and 1.28 TeV, respectively, in $\emph{RL}$ sector of the scharm-stop flavour-violating interaction. We show the changes in marginalised entropy with free CMSSM parameters in the NMFV scenario for the $\emph{LR}$ and $\emph{RL}$ sectors in Figures~\ref{fig:7} and~\ref{fig:8}, respectively. Although the values of free CMSSM parameters, namely unified scalar mass, unified gaugino mass, common trilinear coupling and tan$\beta$ in the CMSSM base model would be 5.99 TeV, 3.58 TeV, $-$6.92 TeV, and 36.8, respectively, as discussed in Ref.~\cite{Gupta:2020whs}. The value of unified scalar and unified gaugino masses decreases to 4.30 TeV and 2.32 TeV, respectively, in $\emph{LR}$ sector due to the implication of flavour-violating interaction. The same consequence is observed in values of trilinear coupling and tan$\beta$, which reduce to $-$4.96 TeV and 22.8, respectively, in the $\emph{LR}$ sector. Similarly, the value of unified scalar mass decreases to 4.16 TeV and unified gaugino mass approaches 3.89 TeV in the $\emph{RL}$ sector. The values of trilinear coupling and tan$\beta$ decline to $-$4.10 TeV and 19.4, respectively, in the $\emph{RL}$ sector. The preferable values of our study can be viewed in Table~\ref{tab:table2} for the $\emph{LR}$ and $\emph{RL}$ sectors of the scharm-stop flavour-violating interaction.

\section{Conclusions}
%In Figure no favoured region is seen particularly in the A0 cattributed to the phenomenon of FV supplementing the
%mass of the Higgs allows for the likelihood of other regions to increase relatively.

In this study, we have explored the effect of flavour-violation of the top-quark with the charm-quark in the context of CMSSM using the information entropy of the Higgs-Boson. In particular, we have performed a detailed random scan over the CMSSM for both cases of $\emph{LR}$ and $\emph{RL}$ flavour-violating couplings of top-quark with charm-quark. For our investigations, we first construct the information entropy of the Higgs-Boson using the branching fraction of its various decay modes over a wide range of CMSSM parameter space for both flavour-violating couplings $\delta_{ct}^{LR}$ and $\delta_{ct}^{RL}$. The information entropy varies with Higgs mass, $m_h$, as shown in Figure~\ref{fig:1}. This clearly reflects that the Higgs mass corresponding to the maximum entropy is in good agreement with the Higgs mass discovered at the LHC~\cite{Aad:2015zhl}. Subsequently, with this persistent approach, we have also explored sparticles masses and the CMSSM parameters by restricting the NMFV parameter space with the experimental constraints obtained from the LEP data, EWPOs, B-Physics, and neutralino dark matter relic density. The findings have been encapsulated in Figures~\ref{fig:2}--\ref{fig:8}. According to our analysis, masses of the neutralino LSP, the lighter chargino and the gluino are expected to be 1.30 TeV, 1.62 TeV, and 4.97 TeV, respectively, corresponding to the case when only $\delta_{ct}^{LR}$ is present. The corresponding estimates for the case with $\delta_{ct}^{RL}$ reflect the masses of the aforementioned sparticles to be 1.28 TeV, 2.24 TeV, and 4.39 TeV, respectively. 
Moreover, the sfermion masses lie in the range from 4.15 TeV to 5.97 TeV for the $\emph{LR}$ sector, whereas 3.92 TeV to 5.81 TeV for the $\emph{RL}$ sector. For $\emph{LR}$ and $\emph{RL}$ sectors, masses of the heavier Higgses turn out to be around 4 TeV or so. The corresponding values of the CMSSM parameters are given by $(m_0$, $m_{1/2}$, $A_{0}$, $tan\beta) = (4.30 {\rm~TeV},  2.32 {\rm~TeV}, -4.96{\rm~ TeV}, 22.8)$ and  $(4.16 {\rm ~TeV}, 3.89~{\rm  TeV}, -4.10~{\rm TeV}, 19.4)$ for the $\emph{LR}$ and $\emph{RL}$ sectors, respectively, corresponding to the maximum entropy. The associated values of $\delta^{LR}_{ct}$ and $\delta^{RL}_{ct}$ are expected to be about 0.037 and 0.039, respectively, for the two cases mentioned above. The values of $m_0$, $m_{1/2}$, $A_{0}$, $tan\beta$ in the case of flavour-conserving CMSSM are expected to be 5.99 TeV, 3.58 TeV, $-$6.92 TeV, and 36.8 according to our earlier study using the same technique in~Ref.~\cite{Gupta:2020whs}. This clearly shows that flavour-violating interactions reduce the SUSY breaking scale considerably, thereby promoting relatively lighter sparticles masses. This also suggests that perchance implementing this effect at the LHC, the signatures of SUSY will certainly be evident in the future.

%%%%%%%%%%%%%%%%%%%%%%%%%%%%%%%%%%%%%%%%%%%%%%%%%
   \begin{table}[h!]
  \begin{center}
    \scriptsize
    \begin{tabular}{cccc|cccc} 
      \hline
      \hline
      \multirow{2}{*}{\textbf{Parameter}}& \multicolumn{3}{c}{\textbf{LR}}&          \multicolumn{3}{c}{\textbf{RL}}&\\
      \cline{2-7} 
       &\textbf{LEP} & \textbf{$+$PO} & \textbf{$+$DM}& 
        \textbf{LEP} & \textbf{$+$PO} & \textbf{$+$DM}\\
       \hline
      $ m_{0}$ &4.26&4.13&4.30&4.43&4.46&4.16\\
      $ m_{1/2}$  &2.45 &2.40 &2.32&2.40 &2.36 &3.89\\
      $ A_{0}$  &$-$4.54 &$-$4.56 &$-$4.96&$-$4.52&$-$4.49 &$-$4.10\\
      $ tan\beta$  &37.7&38.3 &22.8&38.2&38.6&19.4\\
      $\delta^{ij}_{ct}$ &0.036&0.035&0.037&0.030&0.031&0.039\\
       \hline
      $m_h$ &125.47 &125.34  &125.45 &125.28 &125.36&125.38\\
      \hline
      $m_{H}$  &3.61&3.60&4.40&3.49&3.55&4.01\\
      $m_{A^0}$  &3.60&3.60&4.68 &3.45&3.40&3.99\\
      $m_{H^{\pm}}$ &3.69&3.62&4.38&3.55&3.49&3.82\\
      $ m_{\tilde\chi^{0}_{1}}$ &1.08&1.08&1.30&1.01&1.01&1.28\\
      $ m_{\tilde\chi^{0}_{2}}$   &1.85&1.89&1.91&1.79&1.79&1.94\\
       $ m_{\tilde\chi^{0}_{3}}$  &2.26&2.29&1.98&2.16&2.16&1.88\\
      $ m_{\tilde\chi^{0}_{4}}$  &2.31 &2.26&2.01 &2.16&2.19&3.28\\
      $ m_{\tilde\chi^{\pm}_{1}}$     &1.84&1.89&1.62&1.85&1.84&2.24\\
      $ m_{\tilde\chi^{\pm}_{2}}$  &2.26&2.29&2.02 &2.17&2.15&3.34\\
      $ m_{\tilde{g}}$ &5.05&5.17&4.97 &4.55&4.82&4.39\\
      $ m_{\tilde{q}_L}$  &5.63&5.59&5.97&4.98&4.98&5.81\\
      $ m_{\tilde{q}_R}$   &5.26&5.29&5.90&4.90&4.80&5.64\\
      $ m_{\tilde{b}_1}$   &4.78&4.68&5.10&4.32&4.35&4.84\\
      $ m_{\tilde{b}_2}$  &4.98&4.96&5.31&4.60&4.58&4.92\\
      $ m_{\tilde{t}_1}$  &3.98&4.08&4.39&3.84&3.79&4.02\\
      $ m_{\tilde{t}_2}$ &4.78&4.75&4.75&4.31&4.33&4.78\\
      $ m_{\tilde{l}_L}$   &4.08&4.06&4.67&4.35&4.05&4.32\\
      $ m_{\tilde{l}_R}$ &4.16&4.10&4.37&4.19&3.93&4.20\\
      $ m_{\tilde{\tau}_1}$ &3.82&3.95&4.58&4.09&4.22&4.46\\
       $ m_{\tilde{\tau}_2}$ &3.95&3.88&4.15&3.76&3.83&3.92\\
      \hline
      \hline
    \end{tabular}
    \caption{\sf{The sparticle mass spectrum for maximum marginalised entropy in light of NMFV effects in both $\emph{LR}$ and $\emph{RL}$ sectors, including several experimental constraints. All parameters except $ tan\beta$ and $\delta^{ij}_{ct}$ have mass dimension, while $m_{h}$ is in GeV and masses of sparticles are in TeV.}}
    \label{tab:table2}
 \end{center}
\end{table}
\vspace{2em}
\section*{Acknowledgments}
This work was supported in part by University Grant Commission under a Start-Up
Grant no. F30-377/2017 (BSR). We thank Apurba Tiwari for some helpful discussions. We acknowledge the availing of the DST computational lab facility at the Department of Physics, AMU, Aligarh during the initial stage of the work.

\end{document}